\documentclass[twocolumn,aps,prd]{revtex4}
\newcommand{\be}{\begin{equation}}
\newcommand{\ee}{\end{equation}}
\newcommand{\bea}{\begin{eqnarray}}
\newcommand{\eea}{\end{eqnarray}}

\begin{document}
\title{Higgs Masses in the Minimal SUSY SO(10) GUT}
\author{Takeshi Fukuyama$^{\dagger}$} 
\author{Amon Ilakovac$^{\ddagger}$} 
\author{Tatsuru Kikuchi$^{\dagger}$} 
\author{Stjepan Meljanac$^{\star}$} 
\author{Nobuchika Okada$^{\diamond}$}
\affiliation{$^{\dagger}$ 
Department of Physics, Ritsumeikan University, 
Kusatsu, Shiga, 525-8577, Japan}
\affiliation{$^{\ddagger}$
University of Zagreb, Department of Physics, 
P.O. Box 331, Bijeni\v cka cesta 32, HR-10002 Zagreb, Croatia}
\affiliation{$^\star$
Institut Rudjer Bo\v skovi\'c, 
Bijeni\v cka cesta 54, P.O. Box 180, HR-10002 Zagreb, Croatia}
\affiliation{$^\diamond$
Theory Group, KEK, Oho 1-1, Tsukuba, Ibaraki 305-0801, Japan}
\date{\today}
\begin{abstract}
We explicitly show that minimal SUSY $SO(10)$ Higgs-Higgsino mass matrices evaluated by 
various groups are mutually consistent and correct. We 
comment on the corresponding results of other authors. 
We construct one-to-one mappings of our approach to the approaches of 
other authors. 

\end{abstract}
\maketitle
\section{Introduction}
There is a large interest in the minimal supersymmetric $SO(10)$ GUT 
\cite{MinSO10,bm93,fo02,bsv03} concerning neutrino masses 
\cite{bm93, fo02, bsv03}, lepton-flavor violation processes \cite{fko03}, 
and proton decay \cite{fikmo0401,fikmo0406}. To study the proton decay 
lifetime it is important to know Higgs-Higgsino masses which were analyzed 
in Refs. \cite{fikmo0401,fikmo0406,fikmo0405,senj,aul0204,aul0405}. 
However, there are apparently different results for corresponding mass matrices 
\cite{fikmo0401,fikmo0406,fikmo0405,senj,aul0204,aul0405}
for $SO(10)\to G_{321}$ symmetry breaking 
with unbroken supersymmetry 
($G_{321}\equiv SU(3)_c\times SU(2)_L\times U(1)_Y$ is the standard model gauge group). 

In order to prove that the mass matrices 
in Ref. \cite{fikmo0401} are 
correct, we present a set of universal consistency checks 
that the mass matrices must satisfy in our approach. These are the trace of the total 
Higgs mass matrix, the hermiticity condition of the matrix of 
Clebsch-Gordan (CG)-coefficients
in each mass matrix and the higher symmetry group checks 
containing the standard model gauge group
$G_{321}$. Further, finding explicit one-to-one correspondences between 
the results of Refs. \cite{fikmo0401,fikmo0406,fikmo0405,senj,aul0204,aul0405,aul0501},
we prove the consistencies between the approaches considered.\\

\section{Short summary of the minimal SUSY $SO(10)$ GUT}

The Yukawa sector of the minimal SUSY $SO(10)$ GUT has 
couplings of each generation of the matter 
multiplet with only the ${\bf 10}$ and ${\bf \overline{126}}$  Higgs multiplets.
The Higgs sector contains ${\bf 10}\equiv H$, 
${\bf \overline{126}}\equiv \overline{\Delta}$, 
 ${\bf 126}\equiv \Delta$ and ${\bf 210}\equiv \Phi$ multiplets.
 The last two multiplets are necessary to achieve the correct $SO(10)\to G_{321}$ breaking.
The Higgs superpotential reads \cite{fikmo0401}
\bea
\label{W}
W&=& m_1 \Phi^2 + m_2 \Delta \overline{\Delta}
+ m_3 H^2
\nonumber\\
&+& \lambda_1 \Phi^3 + \lambda_2 \Phi \Delta \overline{\Delta}
+ \lambda_3 \Phi \Delta H + \lambda_4 \Phi \overline{\Delta} H.
\label{lee}
\eea
We are interested in symmetry breaking $SO(10)\to G_{321}$. 
The $G_{321}$ invariant VEVs are
\bea
\label{321VEVs}
\langle \Phi \rangle &=& \sum_{i=1}^3 \phi_i \widehat{\phi}_i; \;\;
\langle \Delta \rangle \ =\ v_R\widehat{v_R};\;\;
\langle \overline{\Delta} \rangle \ =\ \overline{v_R}\widehat{\overline{v_R}};
\eea
where $\phi_i,\ i=1,2,3$, $v_R$ and $\overline{v_R}$ 
are complex VEV variables and 
$\widehat{\phi}_i,\ i=1,2,3$, $\widehat{v_R}$ and 
$\widehat{\overline{v_R}}$ are unit $G_{321}$ invariant
vectors, satisfying 
$\widehat{\phi}_i\widehat{\phi}_j = \delta_{ij}$, 
$\widehat{v_R}\widehat{\overline{v_R}}=1$,
$\widehat{v_R}^2={\widehat{\overline{v_R}}}^2=0$, 
described in $Y$ diagonal basis \cite{fikmo0401,fikmo0405} as
\bea
\widehat{v_R}
&=& \frac{1}{\sqrt{120}}(24680),\quad
\widehat{\overline{v_R}}
\ =\
\frac{1}{\sqrt{120}}(13579), 
\nonumber\\
\widehat{\phi}_1 &=& 
-\frac{1}{\sqrt{24}}(1234),
\nonumber\\
\widehat{\phi}_2 &=& 
-\frac{1}{\sqrt{72}}(5678+5690+7890),\;
\nonumber\\
\widehat{\phi}_3 &=& 
-\frac{1}{12}([12+34][56+78+90]).
\label{singlets}
\eea
Inserting the VEVs (\ref{321VEVs}) into the superpotential (\ref{W}), 
one obtains
\bea
\label{VEVW}
\langle W \rangle &=&
m_{1} \left[\phi_1^2 + \phi_2^2 +\phi_3^2 \right]
+ m_{2} v_R \overline{v_R} 
\nonumber\\
&+& 
\lambda_{1} \left[ \phi_2^3 \, \frac{1}{9 \sqrt{2}}
+ 3 \,\phi_1 \phi_3^2 \, \frac{1}{6 \sqrt{6}}
+ 3 \,\phi_2 \phi_3^2 \, \frac{1}{9 \sqrt{2}} \right] 
\nonumber\\
&+& \lambda_{2} \left[\phi_1 \, \frac{1}{10 \sqrt{6}}
+ \phi_2 \, \frac{1}{10 \sqrt{2}}
+ \phi_3 \, \frac{1}{10} \right]
v_R \overline{v_R} ,
\eea
which determines the VEV equations,
\bea
\label{VEVeqs}
0 &=&
2 m_1\phi_1
+\frac{\lambda_1 \phi_3^2}{2\sqrt{6}}
+\frac{\lambda_2 v_R \overline{v_R}}{10\sqrt{6}} ,
\nonumber \\
0 &=&
2 m_1 \phi_2
+\frac{\lambda_1\phi_2^2}{3\sqrt{2}}
+\frac{\lambda_1\phi_3^2}{3\sqrt{2}}
+\frac{\lambda_2 v_R\overline{v_R}}{10\sqrt{2}} ,
\nonumber \\
0 &=&
2 m_1 \phi_3
+\frac{\lambda_1\phi_1 \phi_3}{\sqrt{6}}
+\frac{\sqrt{2}\lambda_1 \phi_2 \phi_3}{3}
+\frac{\lambda_2 v_R\overline{v_R}}{10} ,
\nonumber \\
0 &=&
v_R\overline{v_R}
\left[
m_2
+\frac{\lambda_2\phi_1}{10\sqrt{6}}
+\frac{\lambda_2\phi_2}{10\sqrt{2}}
+\frac{\lambda_2\phi_3}{10} 
\right] .
\eea 

In the following we will assume that $|v_R|\ =\ |\overline{v_R}|$ \cite{fikmo0401,senj}.

For $v_R\ =\ 0$ the solutions of the VEV equations (\ref{VEVeqs}) 
are vacuum minima with $SU(5)\times U(1)$, $SU(5)\times U(1)^{\rm flipped}$,
$G_{3221}$ and $G_{3211}$ symmetry. 

For $v_R\ \neq\ 0$ the VEV equations 
(\ref{VEVeqs}) lead to the fourth-order equation in $\phi_1$ (or $\phi_2$ 
or $\phi_3$). One of the solutions of that equation corresponds to the 
$SU(5)$ symmetry, while the remaining three have $G_{321}$ symmetry
\cite{fikmo0401}. 

The $SU(5)$ solution is given by 
\bea
\phi_1 &=& - \frac{\sqrt{6}m_2}{\lambda_2}, \;
\phi_2 \ =\ - \frac{3\sqrt{2}m_2}{\lambda_2}, \;
\phi_3 \ =\ - \frac{6m_2}{\lambda_2}, 
\nonumber\\
v_R\overline{v_R} & =& 
 \frac{60 m_2^2}{\lambda_2^2}
 \left[2\left(\frac{m_1}{m_2}\right) - 3\left(\frac{\lambda_1}{\lambda_2}\right)\right] \;.
\label{SU5sol}
\eea

\section{Higgs mass matrices}

The mass matrices are defined as
\be
{\cal M}_{ij} \ =\ \frac{\partial^2 W}{\partial 
 \varphi_i \partial \overline{\varphi}_j} \Bigg|_{VEV}.
\ee
where $\varphi_i$ represents any $G_{321}$ multiplet.
We point out that the physical masses squared are
eigenvalues of ${\cal M}^\dagger {\cal M}$ and ${\cal M} {\cal M}^\dagger$ matrices.

The $G_{321}$ mass matrices \cite{fikmo0401,fikmo0405} that we 
use here are given in Ref. \cite{fikmo0401} (see formulas (4.1) -- (4.5), (5.3), (6.4) 
and Tables I and II). Phenomenologically the most interesting doublet and triplet mass matrices are 
given by equations (5.3) and (6.4) in Ref. \cite{fikmo0401}, respectively.

\section{Consistency checks}

In Ref. \cite{fikmo0405} a detailed explanation of a method for calculation 
of the above matrices is given, and all possible consistency 
checks are briefly explained.

There are three main consistency checks. 

The first is that the trace of the total
Higgs mass matrix does not depend on the coupling constants $\lambda_i,\ 
i=1,2,3,4$. It depends only on mass parameters $m_i,\ i=1,2,3$ and the 
dimensions of the corresponding $SO(10)$ representations. The sum rule for the
Higgs-Higgsino mass matrices is
\be
{\rm Tr}{\cal M}\ =\ 
2m_1\times 210  +  m_2\times 252 +  2m_3\times 10 \;.
\ee

The second is that the CG-coefficients in all mass matrices satisfy 
hermiticity property.

The mass sum rule and the hermiticity property for the 
mass matrices are easily verified for the results for the mass matrices
given in Ref. \cite{fikmo0401}.

The third and main check is the $SU(5)$ check briefly 
described in the paper \cite{fikmo0405}.
Here we explicitly prove
that mass matrices in Ref. \cite{fikmo0401} satisfy this highly nontrivial test. 

Let's insert the SU(5) solution (\ref{SU5sol}) 
into mass matrices in Ref. \cite{fikmo0401} for general mass 
parameters $m_i,\ i=1,2,3$ and coupling constants $\lambda_i,\ i=1,2,3,4$.
Note that ${\bf 10}$, ${\bf \overline{126}}$, ${\bf 126}$ and ${\bf 210}$
decompose under the $SU(5)$ symmetry as
\bea
{\bf 10} &=& {\bf 5} + {\bf \overline{5}} ,
\nonumber\\
{\bf 126} &=& {\bf 1} + {\bf \overline{5}} + {\bf 10} 
            + {\bf \overline{15}} + {\bf 45} + {\bf \overline{50}} ,
\nonumber\\
{\bf \overline{126}} &=& {\bf 1} + {\bf 5} + {\bf \overline{10}}
            + {\bf 15} + {\bf \overline{45}} + {\bf 50} ,
\nonumber\\
{\bf 210} &=& {\bf 1} + {\bf 5} + {\bf \overline{5}}   
           +  {\bf 10} + {\bf \overline{10}} +{\bf 24} 
+ {\bf 40} + {\bf \overline{40}} + {\bf 75} .
\nonumber\\
\eea
In total, there are three singlets, three (${\bf 5} + {\bf \overline{5}}$),
two (${\bf 10}$ + ${\bf \overline{10}}$), one (${\bf 15}$ + ${\bf \overline{15}}$),
one ${\bf 24}$, one (${\bf 40}$ + ${\bf \overline{40}}$),
one (${\bf 45}$ + ${\bf \overline{45}}$), one (${\bf 50}$ + ${\bf \overline{50}}$) 
and one ${\bf 75}$. All together there are $14$ $SU(5)$ representations
which can form mass terms. 
  
The corresponding mass-matrix eigenvalues 
are $m_1^G({\bf 1})=m_1^G({\bf 10})=0$,
corresponding to 21 would-be Goldstone modes, 
$m_{2,3}({\bf 1})$, $m_{1,2,3}({\bf 5})$, $m_{2}({\bf 10})$,
$m_\Delta({\bf 15})$, $m_\phi({\bf 24})$, 
$m_\phi({\bf 40})$, $m_\Delta({\bf 45})$, $m_\Delta({\bf 50})$ 
and $m_\phi({\bf 75})$.
Therefore, the $SU(5)$ decomposition of
${\bf 10}$, ${\bf \overline{126}}$, ${\bf 126}$ and ${\bf 210}$ implies there are at
most 13 different mass-matrix eigenvalues.

We found the $SU(5)$ mass-matrix eigenvalues analytically. 
These eigenvalues are obtained diagonalizing
the $26$ mass matrices corresponding to
the $26$ $G_{321}$ multiplets contained in 
${\bf 10}$, ${\bf \overline{126}}$, ${\bf 126}$ and ${\bf 210}$.
The results are given in Table \ref{mE5G321}. In Table \ref{mE5G321} 
the mass-matrix eigenvalues read 
\bea
m_\Delta({\bf 50}) &=& \frac{6}{5} m_2, 
\qquad 
m_\Delta({\bf 45})\ =\ m_2, 
\nonumber\\
\qquad
m_\Delta({\bf 15}) &=& \frac{4}{5} m_2, 
\nonumber\\
m_2({\bf 10}) &=& 2m_1 - 3m_2\frac{\lambda_1}{\lambda_2} + \frac{3}{5} m_2, 
\nonumber\\
m_\phi({\bf 75}) &=& 2m_1 + 2m_2\frac{\lambda_1}{\lambda_2}, 
\qquad
m_\phi({\bf 40})\ =\ 2m_1, 
\nonumber\\
m_\phi({\bf 24}) &=& 2m_1 - m_2\frac{\lambda_1}{\lambda_2},
\nonumber\\
m_{2,3}({\bf 1}) &=& m_1 - 3m_2\frac{\lambda_1}{\lambda_2}
\nonumber\\
\pm\ \bigg[
  \bigg(m_1  \mbox{\hspace{-.5em}}
          &-& \left.\left. \mbox{\hspace{-.7em}} 3m_2\frac{\lambda_1}{\lambda_2}\right)^2
  +4m_1m_2-6m_2^2\frac{\lambda_1}{\lambda_2}
   \right]^{\frac{1}{2}}\;.
\nonumber\\
\eea
The remaining three mass-matrix eigenvalues $m_{1,2,3}({\bf 5})$ 
are solutions of the following cubic equation
\bea
0
&=&
x^3
-x^2 
 \left[
    2 m_1 + 2 m_3 + \frac{3 m_2}{5} - \frac{6 \lambda_1 m_2}{\lambda_2}
 \right]
\nonumber\\
&-&
  x
 \left[
   \frac{36 \lambda_3 \lambda_4 m_2^2}{5 \lambda_2^2}
 - \frac{36 \lambda_1 \lambda_3 \lambda_4 m_2^2}{\lambda_2^3}
 \right.
+
   \frac{9 \lambda_1 m_2^2}{5 \lambda_2}
\nonumber\\
&+&
   \frac{24 \lambda_3 \lambda_4 m_1 m_2}{\lambda_2^2}
 + \left. \frac{12 \lambda_1 m_3 m_2}{\lambda_2}
 - \frac{6 m_3 m_2}{5}
 - 4 m_1 m_3 
 \right]
\nonumber\\
&-&
 \left[
\frac {108 \lambda_1 \lambda_3 \lambda_4 m_2^3}{\lambda_2^3}
- \frac{288 \lambda_3 \lambda_4 m_1 m_2^2}{5 \lambda_2^2}
- \frac{18 \lambda_1 m_3 m_2^2}{5 \lambda_2}
 \right]\;.
\nonumber\\
\eea
\begin{table*}
\caption{$SU(5)$ mass-matrix eigenvalues obtained from $G_{321}$ mass matrices}
\label{mE5G321}
\begin{center}
\begin{tabular}{lclclc|c|}
$G_{321}$ & $SU(5)$ mass-matrix eigenvalues & $G_{321}$ & 
$SU(5)$ mass-matrix eigenvalues \\
$({\bf 3,2},-\frac{5}{6})$     
  & {$m_\phi({\bf 24})$,$m_\phi({\bf 75})$} 
 & $\left({\bf 3,2},\frac{7}{6} \right)$
  & {$m_\Delta({\bf 45})$,$m_\Delta({\bf 50})$}
\\
$({\bf 3,2},\frac{1}{6})$      
  & {$m_1^G({\bf 10})$,$m_2({\bf 10})$,$m_\phi({\bf 40})$,$m_\Delta({\bf 15})$} 
 & $\left({\bf \overline{3},1},\frac{4}{3} \right)$
  & $m_\Delta({\bf 45})$
\\
$({\bf 3,1},\frac{2}{3})$      
  & {$m_1^G({\bf 10})$,$m_2({\bf 10})$,$m_\phi({\bf 40})$}  
 & $\left({\bf 1,3},1 \right)$
  & $m_\Delta({\bf 15})$
\\
$({\bf 1,1},1)$                
  & {$m_1^G({\bf 10})$,$m_2({\bf 10})$} 
 & $\left({\bf 1,1},2 \right)$
  & $m_\Delta({\bf 50})$
\\
$({\bf 1,1},0)$                
  & {$m_1^G({\bf 1})$,$m_2({\bf 1})$,$m_3({\bf 1})$,$m_\phi({\bf 24})$,$m_\phi({\bf 75})$}  
 & $\left({\bf 8,3},0 \right)$
  & $m_\phi({\bf 75})$
\\
$({\bf 1,2},\frac{1}{2})_D$    
  & {$m_\Delta({\bf 45})$,$m_{1}({\bf 5})$,$m_{2}({\bf 5})$,$m_{3}({\bf 5})$} 
 & $\left({\bf 8,1},1 \right)$ 
  & $m_\phi({\bf 40})$
\\
$({\bf 3,1},-\frac{1}{3})_T$   
  & {$m_\Delta({\bf 50})$,$m_\Delta({\bf 45})$,$m_{1}({\bf 5})$,$m_{2}({\bf 5})$,$m_{3}({\bf 5})$}
 & $\left({\bf 8,1},0 \right)$
  & {$m_\phi({\bf 24})$,$m_\phi({\bf 75})$}
\\
$\left({\bf 8,2},\frac{1}{2} \right)$ 
  & {$m_\Delta({\bf 45})$,$m_\Delta({\bf 50})$}
 & $\left({\bf 6,2},\frac{5}{6} \right)$
  & $m_\phi({\bf 75})$  
\\
$\left({\bf 6,3},\frac{1}{3} \right)$
  & {$m_\Delta({\bf 50})$}
 & $\left({\bf 6,2},\frac{1}{6} \right)$
  & $m_\phi({\bf 40})$
\\
$\left({\bf 6,1},\frac{4}{3} \right)$  
  & $m_\Delta({\bf 50})$
 & $\left({\bf \overline{3},3},\frac{2}{3} \right)$
  & $m_\phi({\bf 40})$
\\
$\left({\bf \overline{6},1},\frac{2}{3} \right)$
  & $m_\Delta({\bf 15})$ 
 & $\left({\bf 3,1},\frac{5}{3} \right)$
  & $m_\phi({\bf 75})$
\\
$\left({\bf 6,1},\frac{1}{3} \right)$
  & $m_\Delta({\bf 45})$
 & $\left({\bf 1,3},0 \right)$
  & $m_\phi({\bf 24})$
\\ 
$\left({\bf \overline{3},3},\frac{1}{3} \right)$
  & $m_\Delta({\bf 45})$
 & $\left({\bf 1,2},\frac{3}{2} \right)$
  & $m_\phi({\bf 40})$
\end{tabular}
\end{center}
\end{table*}

We point out that when the $SU(5)$ solution for 
VEVs (\ref{SU5sol}) is inserted in $26$ matrices in Ref. \cite{fikmo0401}
$G_{321}$ mass matrices we obtain $13$ different 
mass-matrix eigenvalues, as predicted 
counting the $SU(5)$ multiplets 
in ${\bf 10}$, 
${\bf \overline{126}}$, ${\bf 126}$ and ${\bf 210}$,
with different mass-matrix eigenvalues.
That is a non-trivial test of the $G_{321}$
mass matrices. 

Moreover, the sum rule for the $SU(5)$ mass-matrix eigenvalues also holds
\bea
{\rm Tr} {\cal M} &=&
[m_{2}({\bf 1})+m_{3}({\bf 1})]
+ [m_{1}({\bf 5}) + m_{2}({\bf 5}) 
\nonumber\\
&+& m_{3}({\bf 5})]\times 10 +
m_2({\bf 10})\times 20 
\nonumber\\
&+& m_\Delta({\bf 15})\times 30 
+ m_\Delta({\bf 45})\times 90
\nonumber\\
&+& m_\Delta({\bf 50})\times 100
+ m_\phi({\bf 24})\times 24 
\nonumber\\
&+& m_\phi({\bf 40})\times 80
+ m_\phi({\bf 75})\times 75
\nonumber\\
&=&
2m_1\times 210 + m_2\times 252 + 2m_3\times 10\;.
\eea

Substitution of the $SU(5)$ solution into $G_{321}$ 
Higgs mass matrices leads, for example, to the following
mass matrices for Higgs doublets $({\bf 1, 2}, \frac{1}{2})$
\begin{eqnarray}
\lefteqn{
M_{\sf doublet} 
\equiv}
\nonumber\\
&&\left(
\begin{array}{cccc}
\begin{array}{c}
2 m_3 \\
0 \\
\frac{6 \lambda_3 m_2}{\sqrt{5} \lambda_2} \\
\frac{2\sqrt{3} \lambda_3 m_2}{\lambda_2} A \\
\end{array}
&\begin{array}{c}
0 \\
m_2 \\
0 \\
0
\end{array}
&\begin{array}{c}
\frac{6 \lambda_3 m_2}{\sqrt{5} \lambda_2} \\
0 \\
\frac{3m_2}{5} \\
- \frac{\sqrt{3} m_2}{\sqrt{5}} A 
\end{array}
&\begin{array}{c}
\frac{2\sqrt{3} \lambda_4 m_2}{\lambda_2} A \\
0 \\
- \frac{\sqrt{3} m_2}{\sqrt{5}} A \\
2 m_1 - \frac{6\lambda_1 m_2}{\lambda_2}
\end{array}
\end{array}
\right)\;,
\nonumber
\\
\label{doubSU5}
\end{eqnarray}
where $A = \left( \frac{2 m_1}{m_2} - \frac{3 \lambda_1}{\lambda_2} \right)^{\frac{1}{2}}$,
and color triplets $({\bf 3, 1}, -\frac{1}{3})$
\begin{eqnarray}
\lefteqn{
M_{\sf triplet} 
\equiv
}
\nonumber
\\
&&
\left(
\begin{array}{ccccc}
%
\begin{array}{c}
2 m_3 \\
M_{21} \\
0 \\
M_{41}\\
M_{51}\\
\end{array}
&
\begin{array}{c}
M_{12} \\
m_2 \\
0 \\
-\frac{m_2}{\sqrt{5}} A \\
-\frac{\sqrt{2} m_2}{5}
\end{array}
&
\begin{array}{c}
0 \\
0 \\
m_2 \\
0 \\
0 \\
\end{array}
&
\begin{array}{c}
M_{14} \\
-\frac{m_2}{\sqrt{5}} A \\
0 \\
M_{44} \\
- \frac{ \sqrt{2} m_2 A}{\sqrt{5}}
\end{array}
&
\begin{array}{c}
M_{51} \\
-\frac{\sqrt{2} m_2}{5} \\
0 \\
- \frac{ \sqrt{2} m_2 A}{\sqrt{5}}  \\
\frac{4 m_2}{5}
\end{array}
\end{array}
\right)\;,
\nonumber
\\
\label{tripSU5}
\end{eqnarray}
. 

where
\bea
M_{12} &=& \frac{2 \sqrt{3} \lambda_4 m_2}{\sqrt{5} \lambda_2},
\qquad
M_{21}\ =\ \frac{2 \sqrt{3} \lambda_3 m_2}{\sqrt{5} \lambda_2},
\nonumber\\
M_{14} &=& 2\sqrt{3} \lambda_4 m_2 A,
\qquad 
M_{41}\ =\ 2\sqrt{3} \lambda_3 m_2 A,
\nonumber\\
M_{44} &=& 2 m_1 -\frac{6 \lambda_1 m_2}{\lambda_2},
\nonumber\\
M_{15} &=& \frac{2 \sqrt{6} \lambda_4 m_2}{\sqrt{5} \lambda_2},
\qquad
M_{51}\ =\ \frac{2 \sqrt{6} \lambda_3 m_2}{\sqrt{5} \lambda_2}.
\eea
Note that
\bea
\label{DTTrDet}
{\rm Tr} \; {\cal M}_{\sf triplet} &=& 
m_\Delta({\bf 50}) + {\rm Tr}\; {\cal M}_{\sf doublet}\;,
\nonumber\\
{\rm det}\; {\cal M}_{\sf triplet} & =& 
m_\Delta({\bf 50}) \times {\rm det}\; {\cal M}_{\sf doublet}\;.
\eea

\section{Equivalence to other approaches}

We show that the results of the Refs. 
\cite{fikmo0401,fikmo0406,fikmo0405,senj,aul0204,aul0405,aul0501} 
are consistent with each other, by giving the unique 
correspondences between the results of different authors. In Ref. 
\cite{aul0501} it was suggested that the issue may be
connected to the different definitions of the
fields. We show explicitly that with the correct field 
identifications the apparently different results are in accord
with each other.

In order to make a contact with the results for the $G_{321}$ mass matrices 
found in Ref. \cite{senj}, we compare the VEV equations 
(6)-(9) of Ref. \cite{senj} and mass matrices given in Table XI and in 
Eqs. (B12)-(B19) of Ref. \cite{senj} with our VEVs equations 
(\ref{VEVeqs}) (see also (3.10)-(3.13) in Ref. \cite{fikmo0401}) 
and mass matrices given in Ref. \cite{fikmo0401} by 
Eqs. (4.1)-(4.5), (5.3), (6.4) and in Tables I and II.
From a comparison of these results in the 
two papers one finds the unique correspondence
between parameters of the two papers, 
\bea
\label{corr}
m_1 &=& m_\phi,\qquad m_2\ =\ m_\Sigma,\qquad m_3\ =\ m_H,
\nonumber\\
\lambda_1 &=& \sqrt{24}\lambda,\qquad \lambda_2\ =\ 10\sqrt{6}\eta,
\nonumber\\
\lambda_3 &=& \sqrt{5}\alpha, \qquad
\lambda_4\ =\ \sqrt{5}\overline{\alpha}, 
\nonumber\\
\phi_1 &=& p, \qquad \phi_2\ =\ \sqrt{3} a,\qquad 
\phi_3\ =\ - \sqrt{6} \omega, 
\nonumber\\
v_R &=& \sigma_B \qquad \overline{v_R}\ =\ \overline{\sigma}_B.
\eea
(Label $B$ is introduced to distingush the quantities
of Ref. \cite{senj} from the equaly named quantities in Ref. 
\cite{aul0405} which will be denoted by label $A$).
Namely, if one performs the above substitution in our 
VEV equations and mass matrices, one gets
the VEV equations as in Ref. \cite{senj}. Also,  
up to the phase redefinitions and 
simultaneous permutations of rows and columns,
the same mass matrices as in Ref. \cite{senj}
are obtained, except for the 
doublet $({\bf 1, 2}, \frac{1}{2})$ mass matrix. 
There is the reverse sign 
in all matrix elements in the fourth row of 
$({\bf 1, 2}, \frac{1}{2})$ mass matrix. 
This difference comes from an arbitrary choice of
phases for states conjugated to each other. In 
our approach the phases of conjugate states are 
chosen to be related by complex conjugation.

Namely, if we multiply our results for the total mass matrix by a diagonal matrix 
of arbitrary phases ${\cal D}$ preserving $G_{321}$ symmetry, 
we obtain matrix ${\cal M}'$, (${\cal M}'= {\cal D}{\cal M}$
or ${\cal M}'= {\cal M}{\cal D}$)
which then spoils all our consistency checks for 
${\cal M}'$, but preserves validity of all our higher symmetry 
checks, except the trace check (\ref{DTTrDet}), for 
$({\cal M}')^\dagger {\cal M}'$ and ${\cal M}' ({\cal M}')^\dagger$ matrices. 
The maximal number of arbitrary phases is equal to the number of $G_{321}$ multiplets.
The matrices ${\cal M}'$ and ${\cal M}$ are physically equvalent.
Hence we agree with \cite{aul0501} that there should be an equivalence
i.e. one-to-one correspondence between all results of all groups.

From the substitution (\ref{corr}) we see that there is one-to-one correspondence 
between VEV equations and mass matrices (up to phases), but the superpotential can be 
identified only after the following rescaling of the ${\bf 210}-\Phi$ and 
${\bf 126}+{\bf \overline{126}}=\Delta+\overline{\Delta}$ fields
\be
\label{corrS}
\Phi\ =\ \frac{\Phi_B}{\sqrt{24}},\qquad
\Delta\ =\ \frac{\Sigma_B}{\sqrt{120}},\qquad
\overline{\Delta}\ =\ \frac{\overline{\Sigma}_B}{\sqrt{120}}, 
\ee
Only after substitutions (\ref{corr}) and the above 
rescalings of the fields (\ref{corrS}) there is
one-to-one correspondence of all our results and results of Ref. \cite{senj}.

The similar equivalence holds between our results and the results of Ref. \cite{aul0405}. But this
is not the correspondence given in Ref. \cite{aul0405} which does not map our $G_{321}$
mass matrices to those of Ref. \cite{aul0405}. The correspondence between our results
and those of Refs. \cite{aul0405,aul0501} is 
\bea
\label{corrA}
m_1 &=& m,\qquad 
m_2\ =\ 2M,\qquad
m_3\ =\ \frac{1}{2} M_H,
\nonumber\\
\lambda_1 &=& \sqrt{24} \lambda,\qquad
\lambda_2\ =\ 20\sqrt{6} \eta,
\nonumber\\
\lambda_3 &=& \sqrt{10} \gamma,\qquad
\lambda_4\ =\ \sqrt{10} \overline{\gamma},
\nonumber\\
\phi_1 &=& p, \qquad
\phi_2\ =\ \sqrt{3} a, \qquad 
\phi_3\ =\ -\sqrt{6} \omega, \qquad
\nonumber\\
v_R &=& \frac{\sigma_A}{\sqrt{2}}, \qquad
\overline{v_R}\ =\ \frac{\overline{\sigma}_A}{\sqrt{2}},
\nonumber\\
\Phi &=& \frac{\Phi_A}{\sqrt{24}}, \qquad
\Delta\ =\ \frac{\Sigma_A}{\sqrt{240}}, \qquad
\overline{\Delta}\ =\ \frac{\overline{\Sigma}_A}{\sqrt{240}} 
\eea

Finally, we have shown that our results are internally consistent and 
correct. By constructing unique mappings from our results to the results of the Refs.
\cite{senj}, and \cite{aul0405,aul0501} we have also shown that the results of 
Refs. \cite{fikmo0401,fikmo0406,fikmo0405,senj,aul0204,aul0405,aul0501} are 
mutually consistent.
The advantages of our method are that it is very simple and necessarily incorporates 
a set of strong consistency checks, proposed in Ref. \cite{fikmo0405} that apply to
the total mass matrix ${\cal M}$. As a
consequence, ${\cal M}^\dagger {\cal M}$ and ${\cal M} {\cal M}^\dagger$ authomatically
satisfy all the higher symmetry tests, except the trace test.
Furthermore, our method can easily be programmed for tensor representations
and can easily be extended to spinor representations. Therefore, it is suitable 
for a broad use in the model
building.

\end{document}